\documentclass[twocolumn,prb,amsmath,amssymb,floatfix,superscriptaddress]{revtex4-2} 
\usepackage{amsmath}
\usepackage{amssymb}
\usepackage{graphicx}
\usepackage{bm}
\usepackage{color}
\usepackage{hyperref}
\usepackage{graphicx}
\usepackage{upgreek}
\usepackage{scalerel}
\usepackage{multirow}
 
 \usepackage[bb=dsserif]{mathalpha}
 
\usepackage{pgffor}
\usepackage{pdfpages}
\usepackage{mathdots}
\usepackage{float}
\usepackage{silence}
\usepackage{physics} 
\usepackage{lscape}   
\usepackage{color, colortbl}
\usepackage[table]{xcolor}
\usepackage{comment}  
\makeatletter
\AtBeginDocument{\let\LS@rot\@undefined}
\makeatother

\begin{document}

\title{Non-Hermitian Josephson junctions with four Majorana zero modes}

\author{Jorge Cayao}
%\email[]{jorge.cayao@physics.uu.se}
\affiliation{Department of Physics and Astronomy, Uppsala University, Box 516, S-751 20 Uppsala, Sweden}
\date{\today}

\author{Masatoshi Sato}
\affiliation{Center for Gravitational Physics and Quantum Information, Yukawa Institute for Theoretical Physics, Kyoto University, Kyoto 606-8502, Japan}

\begin{abstract}
Josephson junctions formed by finite-length topological superconductors host four Majorana zero modes when the phase difference between the superconductors is $\varphi=\pi$ and their length is larger than the Majorana localization length. While this picture is understood in terms of a Hermitian description of isolated junctions, unavoidable transport conditions due to coupling to reservoirs make them  open and ground for non-Hermitian effects that still remain largely unexplored.  In this work, we investigate the impact of non-Hermiticity on  Josephson junctions hosting four Majorana zero modes when they are coupled to normal leads. We demonstrate that, depending on whether inner or outer Majorana zero modes are subjected to non-Hermiticity, Andreev exceptional points  can form between lowest (higher energy) Andreev bound states connected by stable zero real energy lines. We further find that the Andreev exceptional points give rise to strong local and nonlocal spectral weights, thus providing a way for their identification via,  e.g., conductance measurements. Our findings unveil non-Hermiticity for designing non-Hermitian topological phases and for operating  Andreev bound states in Josephson junctions hosting Majorana zero modes.

\end{abstract}
\maketitle

%%%%%%%%%%%%%%%%%%%%%%%%%%%%%%%
% SECTION 1:                 INTRODUCTION                                %
%%%%%%%%%%%%%%%%%%%%%%%%%%%%%%%
\section{Introduction}
Non-Hermitian Josephson junctions have recently been proposed as an outstanding platform for  non-Hermitian topological phenomena   entirely controlled  by the Josephson effect \cite{cayao2023nonhermitian,li2023anomalous}. These ideas have then motivated several studies that have deepened the understanding of non-Hermitian (NH) Josephson junctions (JJs) \cite{shen2024nonhermitian,pino2024thermodynamics,Ohnmacht2024,cayao2024non,li2025EP,PhysRevB.111.064517,ogino2025,solow2025EP,JunjieNHDiode}; see also Ref.\,\cite{JorgeEPs}. A NH JJ represents an open system that naturally forms when coupling a Hermitian JJ to a reservoir and is   described by a NH effective Hamiltonian \cite{cayao2023nonhermitian,beenakker2024josephson}.  Central to NH JJs is the emergence of spectral degeneracies known as exceptional points (EPs) in the Andreev spectrum due to the interplay of non-Hermiticity and the superconducting phase difference between superconductors \cite{cayao2023nonhermitian,li2023anomalous}. At EPs, a pair of Andreev bound states (ABSs) and their respective eigenfunctions coalesce, which represents a NH topological phenomenon protected by particle-hole symmetry PHS${^\dagger}$ \cite{cayao2023nonhermitian}. In JJs formed by two superconductors, the Andreev EPs are connected by stable zero energy lines \cite{cayao2023nonhermitian,li2023anomalous}, while in junctions formed by more superconductors they   appear as   lines and surfaces of Andreev EPs linked by higher dimensional Fermi arcs \cite{Ohnmacht2024,cayao2024non}. Andreev EPs also induce a generic enhancement in the Josephson current  \cite{cayao2023nonhermitian,cayao2024non} and in their spectral signals \cite{cayao2024non},   highlighting  the potential of combining   JJs \cite{RevModPhys.76.411,furusaki1991dc,FukayaJPCM2025} and NH topology \cite{PhysRevX.9.041015,OS23}.

The impact of NH topology on JJs has been    addressed so far in systems with conventional spin-singlet $s$-wave superconductors and in some cases also with unconventional equal spin-triplet (or spinless) $p$-wave superconductors \cite{cayao2023nonhermitian,JorgeEPs,pino2024thermodynamics,cayao2024non,li2025EP}. While the physics of NH JJs with conventional superconductors is by now well understood,  there are still open questions on NH JJs formed by spinless $p$-wave superconductors. Of special relevance is that Hermitian topology in spinless $p$-wave superconductors predicts the formation of boundary states known as Majorana zero modes (MZMs) \cite{tanaka2011symmetry,sato2016majorana,sato2017topological,tanaka2024theory}, with four MZMs in Josephson junctions of finite length \cite{PhysRevLett.108.257001,PhysRevB.91.024514,PhysRevB.96.205425,PhysRevB.104.L020501,PhysRevB.109.L081405,cayao2018andreev,79tj-c3y4}. However, it remains poorly understood how these Hermitian MZMs  respond to non-Hermiticity in JJs and how this response enables the realization of non-Hermitian topological phases with robust signatures. 

\begin{figure}[!b]
\centering
\includegraphics[width=0.45\textwidth]{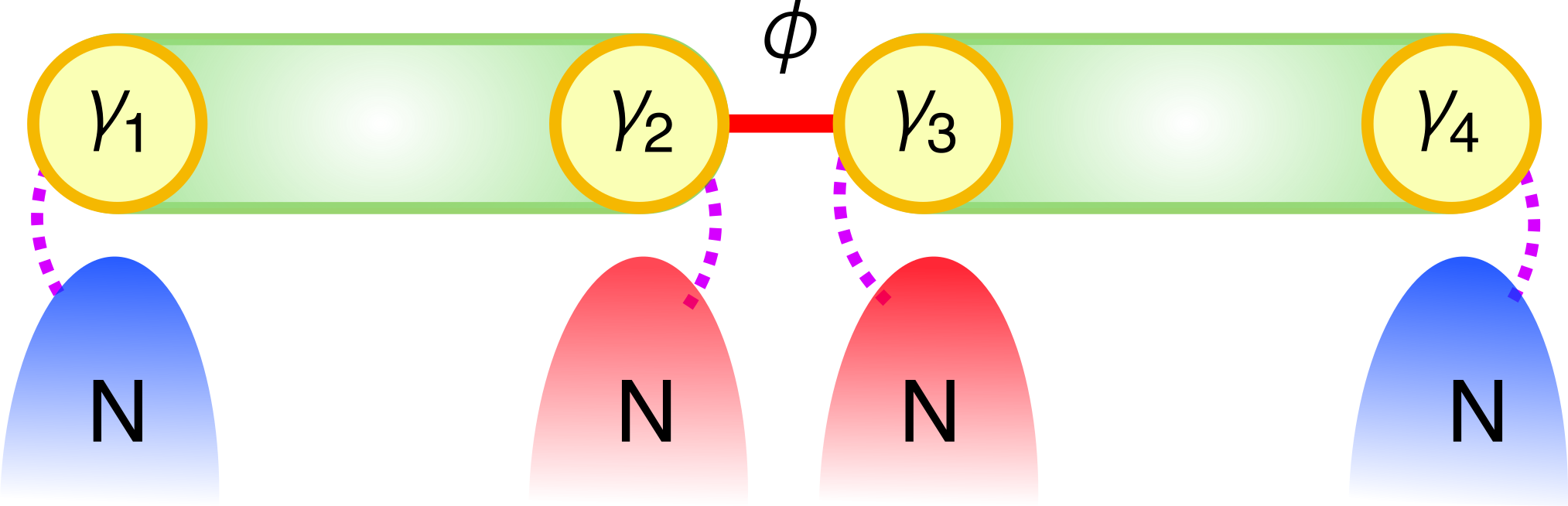}
\caption{A NH Josephson junction with four MZMs ($\gamma_{i}$), where   MZMs emerging at the end of each topological superconductor (green) is coupled to normal N reservoirs (blue and red).}
\label{Fig0} 
\end{figure}

In this work, we consider   phase-biased JJs with four MZMs and investigate the effect of non-Hermiticity when coupling the MZMs  to distinct normal leads or reservoirs [Fig.\,\ref{Fig0}]. We demonstrate that non-Hermiticity induces Andreev EPs which is entirely controlled by the combined effect of the superconducting phase difference and asymmetric couplings to the leads. In particular, by tuning the applied   non-Hermiticity on either   the inner or outer sides of the JJ, where inner or outer MZMs reside, second order Andreev EPs emerge between the lowest ABSs but, stronger couplings to the leads, also lead to EPs between higher energy ABSs. Interestingly, we show that these Andreev EPs are connected by   zero real energy bands, which, due to the inherent NH topology of EPs \cite{OS23,PhysRevX.9.041015,PhysRevLett.123.066405},  represent an stable NH topological phenomenon akin to bulk Fermi arcs  \cite{doi:10.7566/JPSCP.30.011098,PhysRevB.107.104515,PhysRevB.97.041203,PhysRevB.99.165145,kozii2017}. Furthermore, we find that the Andreev EPs originate strong local and nonlocal spectral signatures, offering a solid way for their detection. Our results thus put forward non-Hermiticity as a powerful mechanism for realizing Andreev EPs in JJs with MZMs and also as a way for manipulating Andreev states.

The remainder of this work is organized as follows. In  Section \ref{section2} we 
discuss the effective NH Hamiltonian for NH JJ containing four MZMs. In Section \ref{section3} we discuss the complex energy spectrum and the emergence of Andreev EPs, while in Section \ref{section4} we discuss their spectral signatures. Our conclusions are presented in Section \ref{section5}.

%%%%%%%%%%%%%%%%%%%%%%%%%%%%%%%
% SECTION 2:                 MODELS AND METHOD               %
%%%%%%%%%%%%%%%%%%%%%%%%%%%%%%%
\section{Non-Hermitian Josephson junctions with   Majorana zero modes}
\label{section2}
\begin{figure}[!t]
\centering
\includegraphics[width=0.45\textwidth]{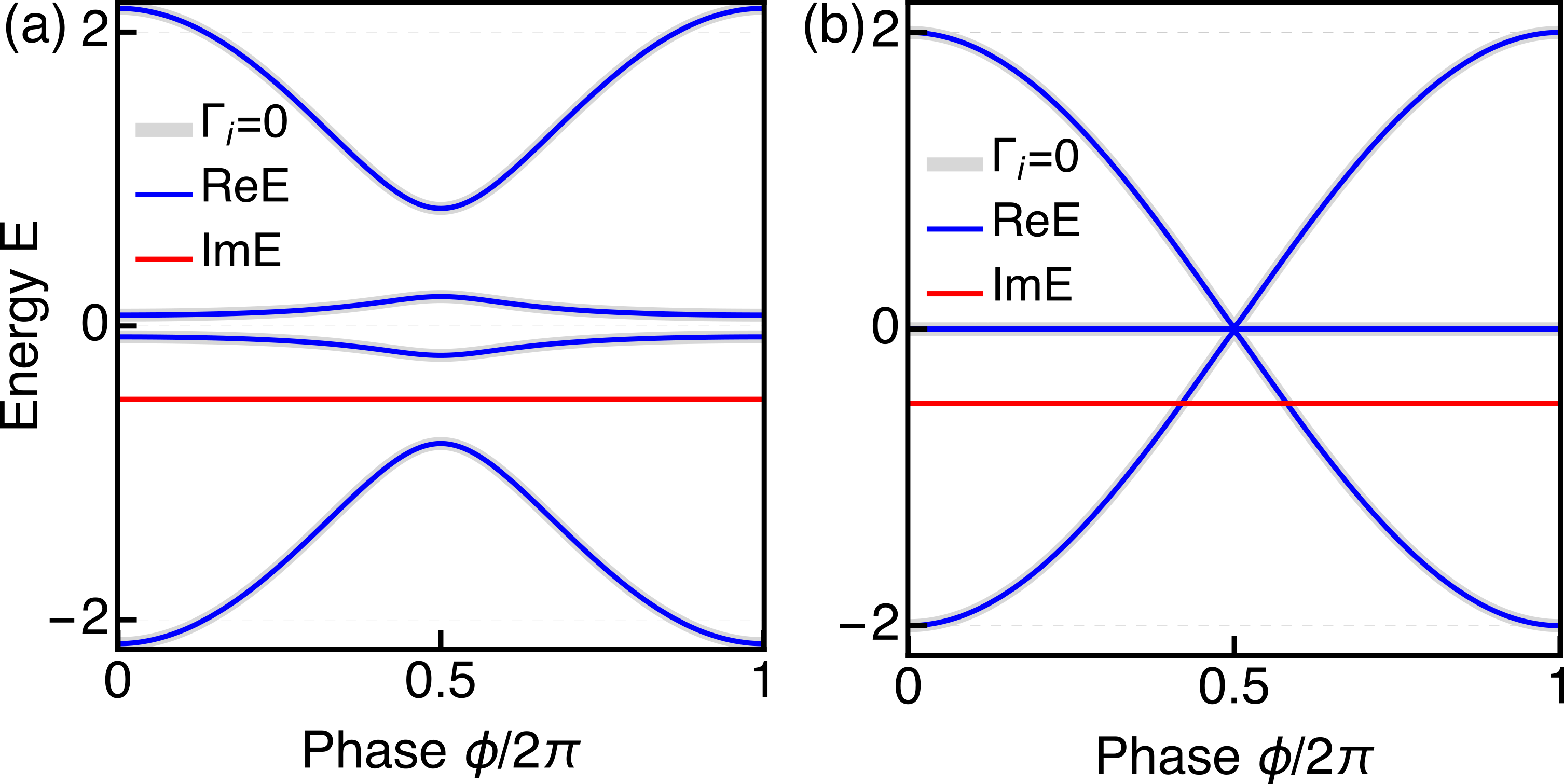}
\caption{Re (blue) and Im (red) parts of the eigenvalues as a function of the superconducting phase difference $\phi$ at finite but equal non-Hermiticity $\Gamma_{i}=0.5$ for  (a) $t'=0.3$, $t=0.5$ and  (b) $t'=0.01$, $t=0.01$. The gray curves below the blue curves correspond to the eigenvalues without non-Hermiticity $\Gamma_{i}=0$. Parameters: $\Gamma_{i}=0.5$, $\bar{\tau}=2$.
}
\label{Fig1} 
\end{figure}

We consider a non-Hermitian Josephson junction hosting four MZMs and coupled to normal (N) reservoirs, as schematically shown in Fig.\,\ref{Fig1}. We model this open Josephson junction using an effective non-Hermitian Hamiltonian given by
\begin{equation}
\label{NHJJEq}
H_{\rm eff}(\phi,\omega)=H_{\rm JJ}+\Sigma^{r}(\omega)\,,
\end{equation}
where $H_{\rm JJ}$ represents the Hermitian Josephson junction while $\Sigma^{r}$ is the retarded self-energy that captures the effect of coupling the MZMs to the normal leads N. More specifically, the Hermitian JJ results from coupling two finite length topological superconductors, which host MZMs at their edges and are modelled by  
\begin{equation}
\label{EqHJJ}
\begin{split}
H_{\rm JJ}(\phi)&=i\bar{\tau}{\rm cos}(\phi/2)\gamma_{2}\gamma_{3}+i(t_{12}\gamma_{2}+t_{13}\gamma_{3})\gamma_{1}\\
&+i(t_{42}\gamma_{2}+t_{43}\gamma_{3})\gamma_{4}+{\rm h.\,c.}
\end{split}.
\end{equation}
Here, $\gamma_{j}$, with $j=1\cdots 4$, denotes the $j$-th   MZM [Fig.\,\ref{Fig0}], $\bar{\tau}=\tau/2$ the coupling between $\gamma_{2}$ and $\gamma_{3}$, $t_{ij}$ the coupling between $\gamma_{i}$ and $\gamma_{j}$ characterizing the spatial overlap between Majorana wavefunctions \cite{PhysRevB.96.205425}, and $\phi$ the superconducting phase difference across the JJ. We further assume that the junction is symmetric and denote $t_{13}=t_{24}\equiv t'$ and $t_{12}=t_{43}\equiv t$. We note that the Hermitian JJ described by Eq.\,(\ref{EqHJJ}) can be derived by directly coupling two Kitaev chains with a superconducting phase difference $\phi$: The coupled Kitaev chains model a Hermitian JJ, which is then transformed into standard Majorana operators \cite{sato2016majorana,sato2017topological,tanaka2024theory} and projected onto the  Majorana operators located at the ends. The $\phi$-dependence in Eq.(\ref{EqHJJ}) captures the 4$\pi$-periodicity and the level crossing between $\gamma_2$ and $\gamma_3$ at $\phi=\pi$.

The retarded  self-energy $\Sigma^{r}$ in Eq.\,(\ref{NHJJEq}) is considered such that each MZM  is coupled to an N reservoir or lead. Moreover, since we are interested in the non-Hermitian effects of the N lead, in the basis $\psi=(\gamma_{1},\gamma_{2},\gamma_{3},\gamma_{4})$, we parametrize its non-Hermitian effect  as
\begin{equation}
\label{selfenergyN}
\Sigma^{r}(\omega=0)=-i\,{\rm diag}(\Gamma_{1},\Gamma_{2},\Gamma_{3},\Gamma_{4})\,.
\end{equation}
In this case, $\Gamma_{j}=\pi |t'_{j}|^{2}\rho_{\rm N}$, with $t'_{j}$ being the hopping between lead and $j$-th MZM and $\rho_{\rm N}$ the surface density of states of the semi-infinite N lead \cite{datta1997electronic,cayao2023exceptional,PhysRevB.105.094502}. For a symmetric junction, we further consider $\Gamma_{2}=\Gamma_{3}$ and $\Gamma_{1}=\Gamma_{4}$, such that the inner  MZMs ($\gamma_{2,3}$) and outer MZMs ($\gamma_{1,4}$) independently are subjected to the same amount of non-Hermiticity. We are interested in exploring the impact of non-Hermiticity on the Josephson junction with four MZMs. For this reason, in what follows, we explore the impact of non-Hermiticity on the energy spectrum and Josephson effect via $\phi$.

Before ending this section, we note that the matrix representation of $H_{\rm eff}(\phi, 0)$ in the basis $\psi=(\gamma_1, \gamma_2, \gamma_3, \gamma_4)$ is purely imaginary, manifesting particle-hole symmetry of class $D^\dagger$ \cite{PhysRevX.9.041015} intrinsic to superconducting systems.
Consequently, the complex energy eigenvalues show the following two distinct behaviors:
While the complex energy eigenvalues always form a pair $(E(\phi), -E^*(\phi))$ when the real part of the eigenvalues is non-zero, they are not when their real part is zero. Then, an EP appears at the transition between these two different behaviors, whose stability is ensured by the above symmetry  \cite{PhysRevLett.123.066405} and we analyze its formation in the next section.

%%%%%%%%%%%%%%%%%%%%%%%%%%%%%%%
% SECTION 3:           Complex energy spectrum                    %
%%%%%%%%%%%%%%%%%%%%%%%%%%%%%%%

\section{Complex energy spectrum: Andreev exceptional points}
\label{section3}
To understand the non-Hermitian effects of the normal leads  on the emergence of four MZMs in the JJs modelled by Eq.\,(\ref{NHJJEq}), we analyze the energy spectrum.  Before going further, we 
first look at the energies in the Hermitian regime ($\Gamma_{i}=0$), which read
\begin{equation}
\label{HRegime4ABSs}
%\begin{split}
E_{\pm}^{\nu}=\pm\Big|\sqrt{t'^{2}+\bar{\tau}^{2}{\rm cos}^{2}(\phi/2)}
-(-1)^{\nu}\sqrt{t^{2}+\bar{\tau}^{2}{\rm cos}^{2}(\phi/2)}\Big|
%\end{split}
\end{equation}
where $\nu$ denotes the lowest ($\nu=0$) and first ($\nu=1$) energy level. To visualize $E^{\nu}_{\pm}$,  the gray curves in Fig.\,\ref{Fig1} show  them as a function of the superconducting phase difference $\phi$ for two values of $t'$ and $t$. We observe the formation of four energy levels, or Andreev bound states (ABSs), which, under general conditions, exhibit a strong dependence on $\phi$. This situation corresponds to $t'<t$, where the lowest energy levels ($E_{\pm}^{0}$) develop a peaked shape around $\phi=\pi$ and a reduced value near to $\phi=0,2\pi$, see Fig.\,\ref{Fig1}(a). Moreover, here, the first excited levels ($E_{\pm}^{1}$) detach from the gap edges at $\phi=0,2\pi$ and follow a cosine-like profile, with a minimum at $\phi=\pi$ where they also anticross with the lowest Andreev levels [Fig.\,\ref{Fig1}(a)].  The size of the anticrossing gap between the lowest and first excited levels can be controlled by the ratio between $t'$ and $t$. It is worth noting that having $t'<t$ is perhaps expected in  a JJ  with finite length superconductors, highlighting their role as indicators of the spatial overlap between Majorana wavefunctions \cite{PhysRevB.110.224510}:  the  spatial overlap between the wavefunctions of $\gamma_{1(3)}$ and $\gamma_{2(4)}$) is very likely to be  stronger than  the overlap between $\gamma_{1(2)}$ and $\gamma_{3(4)}$ \cite{PhysRevB.96.205425}.  In the case when $t'=t$,  the lowest energy levels  acquire zero energy ($E_{\pm}^{0}=0$) and do not disperse with $\phi$, see Fig.\,\ref{Fig1}(b) and Eq.\,(\ref{HRegime4ABSs}). Furthermore, the first excited levels tend to reach zero energy at $\phi=\pi$ as $t,t'\rightarrow0$, as seen in Fig.\,\ref{Fig1}(b), expected to happen in a JJ with very large superconductors. Here, the dispersionless levels $E_{\pm}^{0}$ define a pair of MZMs located at the outer ends of the JJs ($\gamma_{1,4}$), while and additional pair of MZMs emerges at $\phi=\pi$ due to $E_{\pm}^{1}$ and located at the inner sides of the JJ ($\gamma_{2,3}$). Thus, the formation of four MZMs, two outer $\gamma_{1,4}$ and two inner $\gamma_{2,3}$  as well as their hybridization, is fully captured by the Hermitian  Hamiltonian in Eq.\,(\ref{EqHJJ}), which is consistent with the properties of MZMs in more realistic models based on Rashba semiconductors with proximity induced superconductivity \cite{PhysRevLett.108.257001,PhysRevB.86.140504,Cayao2020Oddfrequency,PhysRevB.96.205425,cayao2018andreev,PhysRevLett.123.117001,baldo2023zero,PhysRevB.109.L081405,fksg-x8pr}.

Having discussed the Hermitian MZMs, we now turn our attention to their behavior under non-Hermiticity due to coupling MZMs to normal leads, as modelled by Eq.\,(\ref{NHJJEq}). In general, the non-Hermitian effect of the leads, captured in the retarded self-energy given by Eq.\,(\ref{selfenergyN}), originates from its imaginary nature \cite{JorgeEPs,cayao2023exceptional,PhysRevB.105.094502,PhysRevB.107.104515,avila2019non,cayaominimalkitaev}. This gives rise to an energy spectrum that is complex \cite{Moiseyev,pikulin2012topological,RevModPhys.87.1037}, as expected for open systems \cite{datta1997electronic}.  The real (Re) part represents the physical energy, while the inverse of its imaginary (Im) part corresponds to the lifetime or decay rate \cite{Moiseyev,el2018non,datta1997electronic}.  When the couplings to the normal leads are the same for the four MZMs, $\Gamma_{i}=\Gamma$, we obtain $\tilde{E}_{\pm}^{\nu}=-\Gamma\pm E_{\pm}^{\nu}$, which implies that  all the energy levels acquire the same Im part, while the Re parts have the same value as those in the Hermitian regime, see blue and red curves in Fig.\,\ref{Fig1}(a,b). However, when the couplings $\Gamma_{i}$
acquire distinct values, the expressions for the eigenvalues are more complicated than those shown in Eqs.\,(\ref{HRegime4ABSs}) and  the emergent physics becomes richer. In what follows, we   consider two cases: i)   outer MZMs   coupled to leads ($\Gamma_{1,4}\neq0$ and   $\Gamma_{2,3}=0$), and ii)    inner MZMs coupled to leads ($\Gamma_{2,3}\neq0$ and   $\Gamma_{1,4}=0$).

%%%%%%%%%%%%%%%%%%%%%%%%%%%%%%%%%%%%%
% SUBSECTION 3.2:          Outer MZMs under non-Hermiticity                 %
%%%%%%%%%%%%%%%%%%%%%%%%%%%%%%%%%%%%%

\subsection{Outer MZMs under non-Hermiticity}
We first analyze the case when outer MZMs ($\gamma_{1,4}$) are coupled to the normal leads by $\Gamma_{1,4}\neq0$, while the inner MZMs ($\gamma_{2,3}$) remain uncoupled with $\Gamma_{2,3}=0$. The eigenvalues in this case do not acquire a simple form but they can be found by solving  $\mathcal{D}(\omega,\phi)\equiv{\rm det}(\omega-H_{\rm eff}(\phi,0))=0$, where we have
\begin{equation}
\label{DDOuter14}
\mathcal{D}(\omega,\phi)=\omega^{2}\mathcal{P}(\omega,\phi)+i\omega(\Gamma_{1}+\Gamma_{4})\mathcal{Q}(\omega,\phi)+ \mathcal{R}(\phi),
\end{equation}
with
\begin{equation}
\label{DDOuter14X}
\begin{split}
\mathcal{P}(\omega,\phi)&=\omega^{2}-2(t'^{2}+t^{2})-\Gamma_{1}\Gamma_{4}-\bar{\tau}^{2}{\rm cos}^{2}(\phi/2)\,,\\
\mathcal{Q}(\omega,\phi)&=[\omega^{2}-(t'^{2}+t^{2})-\bar{\tau}^{2}{\rm cos}^{2}(\phi/2)]\,,\\
\mathcal{R}(\phi)&=(t'^{2}-t^{2})^{2}+\Gamma_{1}\Gamma_{4}\bar{\tau}^{2}{\rm cos}^{2}(\phi/2)\,.
\end{split}
\end{equation}
Here and below, we use the notation $H_{\rm eff}(\phi,0)$ to represent its matrix representation.
%Thus, the spectrum is simply given by the poles of the Green's function, $\mathcal{D}(\omega,\phi)\equiv{\rm det}(\omega-H_{\rm eff}(\phi,0))=0$. 
At this point, we can directly see the effect of non-Hermiticity in the coefficient of $\mathcal{Q}$, which is absent in the Hermitian regime $\Gamma_{1,4}=0$.  An analytical expression for the lowest pair of ABSs can be found by solving $\mathcal{D}(\omega,\phi)=0$ with $\mathcal{D}(\omega,\phi)$ up to second order in $\omega$. Then, we obtain
\begin{equation}
\label{ABSEQ7}
\omega_{\pm}(\phi)=-2i\Gamma_{14}\mathcal{A}(\phi)\pm \sqrt{\mathcal{B}(\phi)-4\Gamma_{14}^{2}\mathcal{A}^{2}(\phi)}\,,
\end{equation}
where $\Gamma_{14}=(\Gamma_{1}+\Gamma_{4})/2$, $\mathcal{A}(\phi)=[t'^{2}+t^{2}+\bar{\tau}^{2}{\rm cos}^{2}(\phi/2)]/\mathcal{A}_{0}(\phi)$, $\mathcal{B}(\phi)=[2(t'^{2}-t^{2})^{2}+2\Gamma_{1}\Gamma_{4}\bar{\tau}^{2}{\rm cos}^{2}(\phi/2)]/\mathcal{A}_{0}(\phi)$, $\mathcal{A}_{0}(\phi)=4t'^{2}+4t^{2}+2\Gamma_{1}\Gamma_{4}+2\bar{\tau}^{2}{\rm cos}^{2}(\phi/2)$. Thus, the two ABSs merge into a single value at $\mathcal{B}(\phi)-4\Gamma_{14}^{2}\mathcal{A}^{2}(\phi)=0$, which defines the emergence of second order EPs between outer MZMs at phases $\phi_{\rm EP}$   around $\phi=\pi$. 

\begin{figure}[t]
\centering
\includegraphics[width=0.49\textwidth]{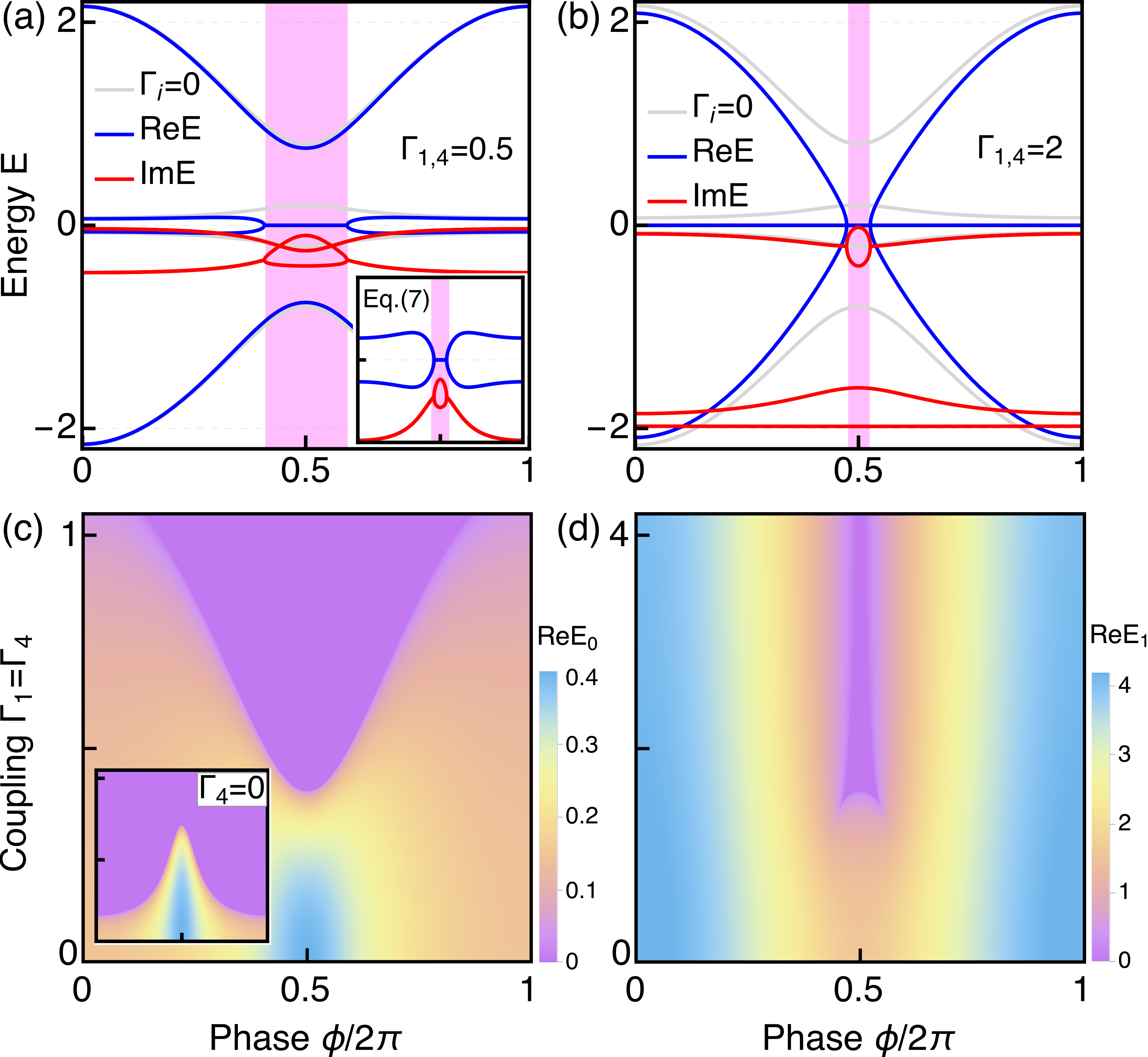}
\caption{(a,b) Re (blue) and Im (red) parts of the eigenvalues as a function of the superconducting phase difference $\phi$ when $\Gamma_{2,3}=0$ for (a) $\Gamma_{1,4}=0.5$ and (b) $\Gamma_{1,4}=2$, in both cases with $t'=0.3$, $t=0.5$. The inset shows the Re and Im energies obtained from Eq.\,(\ref{ABSEQ7}) for the same regime of  (a).  The gray curves correspond to the eigenvalues without non-Hermiticity $\Gamma_{i}=0$. The ends of the shaded magenta regions mark the position of EPs connecting zero  Re energy lines. (c) Re part of the energy difference between the   positive and negative levels closest to zero real energy as a function of $\phi$ and $\Gamma_{1}=\Gamma_{4}$, with the magenta regions indicating the zero Re energy and their ends marking the EPs. The inset in (c) shows the same as in (c) but at $\Gamma_{4}=0$. (d) Same as in (c) but for the Re part of the energy difference between first excited positive and negative levels. Parameters: $\bar{\tau}=2$.
}
\label{Fig2} 
\end{figure}

The formation of EPs given by Eqs.\,(\ref{ABSEQ7}) as a function of $\phi$ is shown in the inset of Fig.\,\ref{Fig2}(a), where we show  $\omega_{\pm}(\phi)$ at $\Gamma_{1,4}=0.5$ for  sizeable finite values of     $t'$ and $t$;  here, the ends of the shaded magenta region   mark the EPs. We have checked that the associated wavefunctions of the pair of ABSs also merge into a single wavefunction at EPs, in line with the definition of EPs where both eigenvalues and eigenvectors coalesce \cite{Heiss_2012,doi:10.1080/00018732.2021.1876991,RevModPhys.93.015005}.  It is worth reminding that $t'$  is the coupling between the leftmost (rightmost) outer MZM ($\gamma_{1 (4)}$) with the right (left) inner MZM ($\gamma_{3 (2)}$), while  $t$ is the coupling between MZMs in the same superconductor; for the NH case with $t',t\rightarrow0$, see subsection \ref{subsectionIIIC}.  Above and below   $\phi_{\rm EP}$, and close to $\phi=0,2\pi$, the Im parts of the two lowest ABSs  are degenerate and also dependent  on $\phi$.  When $\mathcal{B}(\phi)-4\Gamma_{14}^{2}\mathcal{A}^{2}(\phi)<0$, the square root produces only Im parts, giving rise to a zero Re energy line connecting the EPs, indicated by the shaded magenta region in the inset of Fig.\,\ref{Fig2}(a). Here, the Im parts between EPs split following an egg-like profile that tends to approach zero at $\phi=\pi$, see red curves in the inset of Fig.\,\ref{Fig2}(a) at low couplings $\Gamma_{1,4}=0.5$.  The behavior of the ABSs and their EPs revealed by the analytical expressions in Eqs.\,(\ref{ABSEQ7}) is preserved when fully solving  the fourth order polynomial   $\mathcal{D}(\omega,\phi)=0$, as we demonstrate in Fig.\,\ref{Fig2}(a) at $\Gamma_{1,4}=0.5$. It is worth noting that, while the lowest ABSs undergo two EP transitions around $\phi=\pi$, the Re parts of the higher energy ABSs still follow the behavior of the Hermitian regime but acquire  nonzero degenerate Im parts that weakly depend on the phase around $\phi=\pi$.

When the couplings to the leads increase  further, the ABSs closest to zero Re energy acquire zero energy  and become  dispersionless flat bands for any $\phi$, see Fig.\,\ref{Fig2}(b) for  $\Gamma_{2,4}=2$. Still, these flat zero energy Re bands, associated to the outer MZMs, possess finite Im parts which appear at higher negative values.  Moreover, the strong non-Hermiticity affects the higher energy levels, making them to develop EP transitions around $\phi=\pi$, with a zero energy Re line  connecting the formed EPs and their Im parts forming a circular profile very close to zero; see blue and red curves within the shaded magenta region in Fig.\,\ref{Fig2}(b). The dispersionless behavior of the NH ABSs closes to zero energy in Fig.\,\ref{Fig2}(b) resembles the phase dependence of the Hermitian outer MZMs for $t',t\rightarrow0$ [Fig.\,\ref{Fig1}(b)], which is, however, expected in    very long superconductors; here we obtain this physics at sizeable finite values of $t',t$, namely, in shorter superconductors; for the NH case with $t',t\rightarrow0$, see subsection \ref{subsectionIIIC}. The ability to control EPs with zero Re energy lines can be further seen in Fig.\,\ref{Fig2}(c), where we show the Re part of the difference between   energies closest to zero  at $\Gamma_{1}\equiv\Gamma_{4}\neq0$. In this case [Fig.\,\ref{Fig1}(c)], the magenta region reflects the zero Re energy difference and its borders mark the emergence of EPs. We see that EPs and their connecting zero Re energy lines start forming slightly below $\Gamma_{1,4}=0.5$ and, as such couplings increase, the length of the zero Re energy line gets longer with the edges following a V-shaped profile around $\phi=\pi$. However, if one coupling is set to zero, e. g.,  $\Gamma_{4}=0$, EPs can form   around $\phi=0$  and a fully zero Re energy line for any $\phi$ can be achieved, both already at lower couplings than for $\Gamma_{1,4}\neq0$; to see these effects, compare the inset in Fig.\,\ref{Fig1}(c) with Fig.\,\ref{Fig1}(c) itself.  But increasing the amount of non-Hermiticity ($\Gamma_{1,4}$) also affects the first excited levels, whose Re difference can also acquire zero value, which is revealed in Fig.\,\ref{Fig1}(d) by the magenta region appearing slightly below   $\Gamma_{1,4}=2$ around $\phi=\pi$. Thus, depending on the need of MZMs, tailored non-Hermiticity can be used to promote or enhance the presence of either inner or outer MZMs

%%%%%%%%%%%%%%%%%%%%%%%%%%%%%%%%%%%%%
% SUBSECTION 3.2:          Inner MZMs under non-Hermiticity                 %
%%%%%%%%%%%%%%%%%%%%%%%%%%%%%%%%%%%%%

\subsection{Inner MZMs under non-Hermiticity}
\begin{figure}[!t]
\centering
\includegraphics[width=0.49\textwidth]{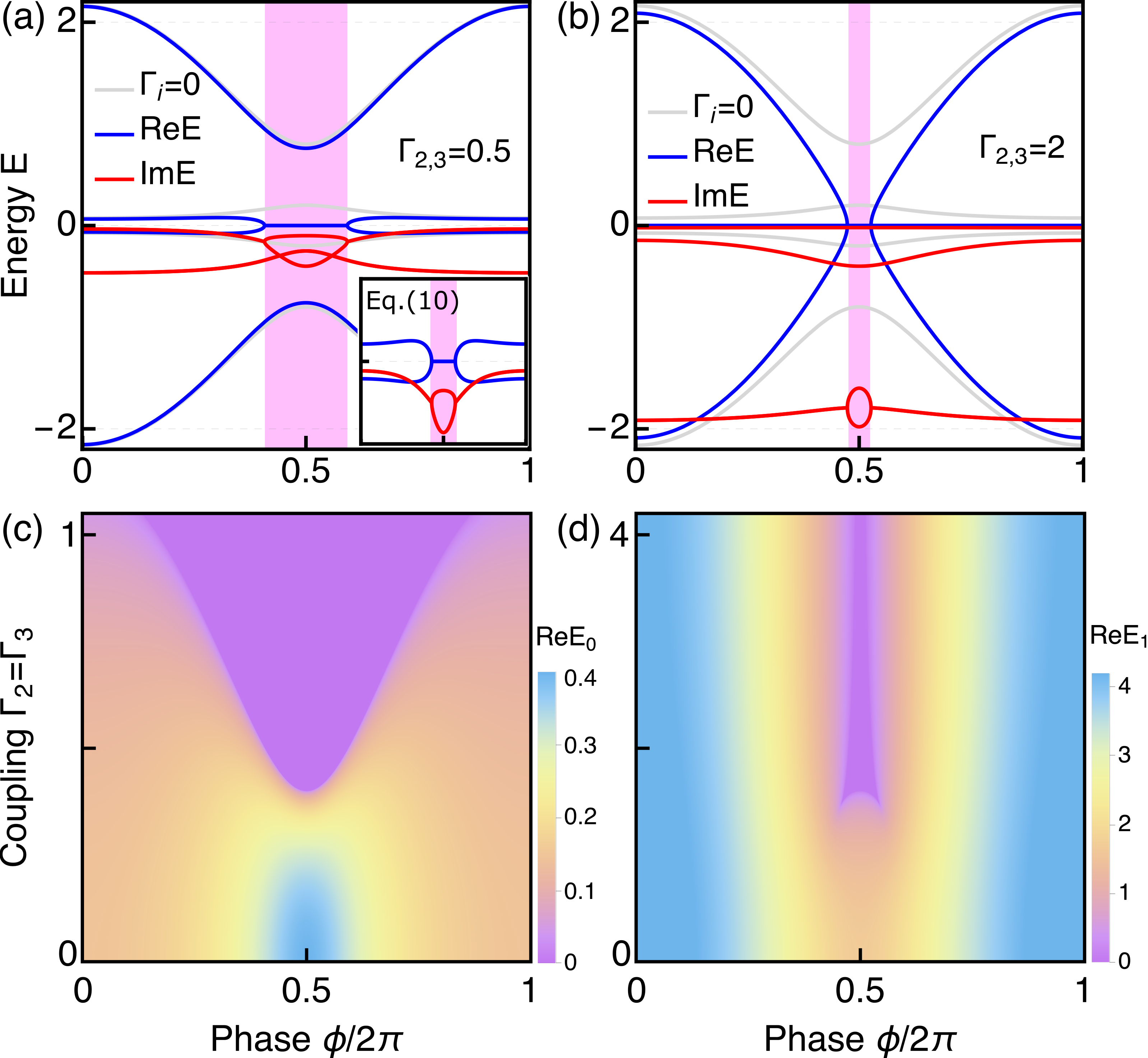}
\caption{(a,b) Re (blue) and Im (red) parts of the eigenvalues as a function of   $\phi$ when $\Gamma_{1,4}=0$ for (a) $\Gamma_{2,3}=0.5$ and (b) $\Gamma_{2,3}=2$, in both cases with $t'=0.3$, $t=0.5$. The gray curves correspond to the eigenvalues without non-Hermiticity $\Gamma_{i}=0$.  The inset shows the Re and Im energies obtained from Eq.\,(\ref{EqInnerMZMsG}) for the same regime of  (a). The ends of the shaded magenta regions mark the position of EPs connecting zero  Re energy lines. (c) Re part of the energy difference between the   positive and negative levels closes to zero as a function of $\phi$ and $\Gamma_{2}=\Gamma_{3}$, with the magenta regions indicating the zero Re energy and their ends marking the EPs.  (d) Same as in (c) but for the Re part of the energy difference between first excited positive and negative levels. Parameters: $\bar{\tau}=2$.}
\label{Fig3} 
\end{figure}

We now consider that only the inner MZMs ($\gamma_{2,3}$) are coupled to the normal leads and hence $\Gamma_{2,3}\neq0$ but $\Gamma_{1,4}=0$. As discussed in the previous subsection, to explore the effect of non-Hermiticity due to $\Gamma_{2,3}\neq0$, we obtain the energy spectrum by finding the poles of the Green's function associated to Eq.\,(\ref{NHJJEq}). This means that we have to solve  $\bar{\mathcal{D}}(\omega,\phi)=0$ for $\omega$, where
\begin{equation}
\label{barD}
\bar{\mathcal{D}}(\omega,\phi)=\omega^{2}\bar{\mathcal{P}}(\omega,\phi)+i\omega(\Gamma_{2}+\Gamma_{3})\bar{\mathcal{Q}}(\omega)+ \bar{\mathcal{R}}\,,
\end{equation}
which is similar to Eq.\,(\ref{DDOuter14}) but $\bar{\mathcal{P}}$, $\bar{\mathcal{Q}}$, and $\bar{\mathcal{R}}$ acquire slightly different  dependences on $\omega$ and $\phi$. In fact, $\bar{\mathcal{P}}$, $\bar{\mathcal{Q}}$, and $\bar{\mathcal{R}}$  are given by 
\begin{equation}
\label{barPQR}
\begin{split}
\bar{\mathcal{P}}(\omega,\phi)&=\omega^{2}-2(t'^{2}+t^{2})-\Gamma_{2}\Gamma_{3}-\bar{\tau}^{2}{\rm cos}^{2}(\phi/2)\,,\\
\bar{\mathcal{Q}}(\omega)&=[\omega^{2}-(t'^{2}+t^{2})]\,,\\
\bar{\mathcal{R}}&=(t'^{2}-t^{2})^{2}\,.
\end{split}
\end{equation}
From these expressions, and comparing them with Eqs.\,(\ref{DDOuter14X}), we clearly see that $\bar{\mathcal{P}}(\omega,\phi)$ remains dependent on $\omega$ and $\phi$ but with the effect of $\Gamma_{2,3}$, $\bar{\mathcal{Q}}(\omega)$ is only dependent on $\omega$ but not on $\phi$, while $\bar{\mathcal{R}}$ does not depend on $\omega$ and $\phi$. We can then analytically obtain the lowest pair of ABSs by solving $\bar{\mathcal{D}}(\omega,\phi)=0$ with $\bar{\mathcal{D}}(\omega,\phi)$ expanded up to second order in $\omega$. Hence, we find
\begin{equation}
\label{EqInnerMZMsG}
\omega_{\pm}(\phi)=-2i\Gamma_{23}\bar{\mathcal{A}}(\phi)\pm\sqrt{\bar{\mathcal{B}}(\phi)-4\Gamma_{23}^{2}\bar{\mathcal{A}}^{2}(\phi)}\,,
\end{equation}
where $\Gamma_{23}=(\Gamma_{2}+\Gamma_{3})/2$, $\bar{\mathcal{A}}(\phi)=[t'^{2}+t^{2}]/\bar{\mathcal{A}}_{0}(\phi)$, $\bar{\mathcal{B}}(\phi)=[2(t'^{2}-t^{2})^{2}]/\bar{\mathcal{A}}_{0}(\phi)$, $\bar{\mathcal{A}}_{0}(\phi)=4t'^{2}+4t^{2}+2\Gamma_{2}\Gamma_{3}+2\bar{\tau}^{2}{\rm cos}^{2}(\phi/2)$. By a direct inspection of Eq.\,(\ref{EqInnerMZMsG}), the pair of ABSs acquire zero Re energy and the same Im part when the square root vanishes, namely, $\bar{\mathcal{B}}(\phi)=4\Gamma_{23}^{2}\bar{\mathcal{A}}^{2}(\phi)$. As in the previous subsection, this condition defines the  values of $\phi$ at which the lowest ABSs merge into a single, which  signals the emergence of Andreev EPs. Moreover, when $\bar{\mathcal{B}}(\phi)<4\Gamma_{23}^{2}\bar{\mathcal{A}}^{2}(\phi)$, the pair of ABSs $\omega_{\pm}(\phi)$ acquire zero Re energy but different Im parts connecting the pair of Andreev EPs. To visualize these Andreev EPs, in    the inset of Fig.\,\ref{Fig3}(a) we present $\omega_{\pm}(\phi)$ from Eqs.\,(\ref{EqInnerMZMsG}) as a function of $\phi$ at low $\Gamma_{2,3}=0.5$ and finite values of $t'$,   $t$, and $\tau$. Here, we clearly see that the Re parts of the ABSs merge into a single zero energy value at phases around   $\phi=\pi$ connected by a zero Re energy line, within which the Im are split; the ends of the magenta shaded region indicates the  EPs.
 We have also verified that the wavefunctions of such ABSs also merge into a single wavefunction, consistent with the definition of EPs where both eigenvalues and eigenvectors coalesce \cite{Heiss_2012,OS23,doi:10.1080/00018732.2021.1876991,RevModPhys.93.015005}. While there is a similarity with the Andreev EPs due to coupling the outer MZMs to normal leads [Fig.\,\ref{Fig2}(a)], the profile of the Im parts when inner MZMs in the inset of Fig.\,\ref{Fig2}(a) develops an   inverted egg-shape profile that does not tend to approach zero at $\phi=\pi$ but instead points towards negative energies, unlike the inset of Fig.\,\ref{Fig2}(a). This   innocent variation originates from the fact that the pair of ABSs in Eq.\,(\ref{EqInnerMZMsG}) characterizes the outer MZMs which are now not coupled to the normal leads since $\Gamma_{2,3}\neq0$ and $\Gamma_{1,4}=0$. The phase dependence of the lowest pair of ABSs shown in the inset of Fig.\,\ref{Fig3}(a) is fully preserved when solving  $\bar{\mathcal{D}}(\omega,\phi)=0$ with $\bar{\mathcal{D}}(\omega,\phi)$ given by Eq.\,(\ref{barD}) without any low-frequency approximation, as we indeed show in Fig.\,\ref{Fig3}(a). Here we also see that the Re parts of the higher energy ABSs remain as in the Hermitian regime (gray curves), but their Im parts acquire the same phase dependence that tends to reach zero at $\phi=\pi$, which is distinct to the case when  outer MZMs are subjected to non-Hermiticity in Fig.\,\ref{Fig2}(a).

By allowing the couplings to the leads $\Gamma_{2,3}$ to take larger values, the lowest and higher ABSs also exhibit important changes, see Fig.\,\ref{Fig3}(b). The Re energies of the lowest pair of ABSs now become zero energy flat bands with $\phi$, with split Im parts where one of them weakly depends on   the phase near $\phi=\pi$, while the other vanishes due to the strong non-Hermiticity. This implies that one of the outer MZMs   always remains at the outer side of the JJ, but the other outer MZM having a finite Im part     decays into the normal lead.  Interestingly, at these values of non-Hermiticity induced by large $\Gamma_{2,3}$, the higher energy ABSs develop EPs around $\phi=\pi$ [Fig.\,\ref{Fig3}(b)], with the Re parts forming zero energy lines that connect EPs while  the Im parts acquiring a split circular profile  between EPs. It is worth noting that the Re and Im parts of the ABSs behave in a similar way as those shown in Fig.\,\ref{Fig2}(b) for non-Hermiticity applied only to the outer MZMs; the only difference is that the Im parts of the higher energy ABSs developing EPs in Fig.\,\ref{Fig3}(b) appear at higher values. This effect is a clear consequence of that the higher energy ABSs form inner MZMs, which are those that are now subjected to a large value of non-Hermiticity via $\Gamma_{2,3}$. The impact of strong non-Hermiticity can be further visualized in Fig.\,\ref{Fig3}(c,d), where we plot the Re part of the energy difference  between  the positive and negative ABSs closes to zero as well as that between higher energy ABSs, both as functions of $\Gamma_{2}=\Gamma_{3}$ and $\phi$. The formation of stable EPs and their connecting zero Re energy lines formed between lowest ABSs remains as non-Hermiticity increases via $\Gamma_{2,3}$ [Fig.\,\ref{Fig3}(c)], and, for very large values of non-Hermiticity, the higher energy ABSs also host stable EPs [Fig.\,\ref{Fig3}(d)]. Thus, by adjusting the amount of non-Hermiticity applied to either inner or outer MZMs, due to their coupling to normal leads, it is possible to  realize EPs and zero Re energy likes akin to bulk Fermi arcs as a unique effect of non-Hermitian topology, which further offers a unique way for controlling the dispersion of ABSs.

%%%%%%%%%%%%%%%%%%%%%%%%%%%%%%%%%%%%%
% SUBSECTION 3.3:        Flat outer MZMs  in the Hermitian regime         %
%%%%%%%%%%%%%%%%%%%%%%%%%%%%%%%%%%%%%

\begin{figure}[!t]
\centering
\includegraphics[width=0.49\textwidth]{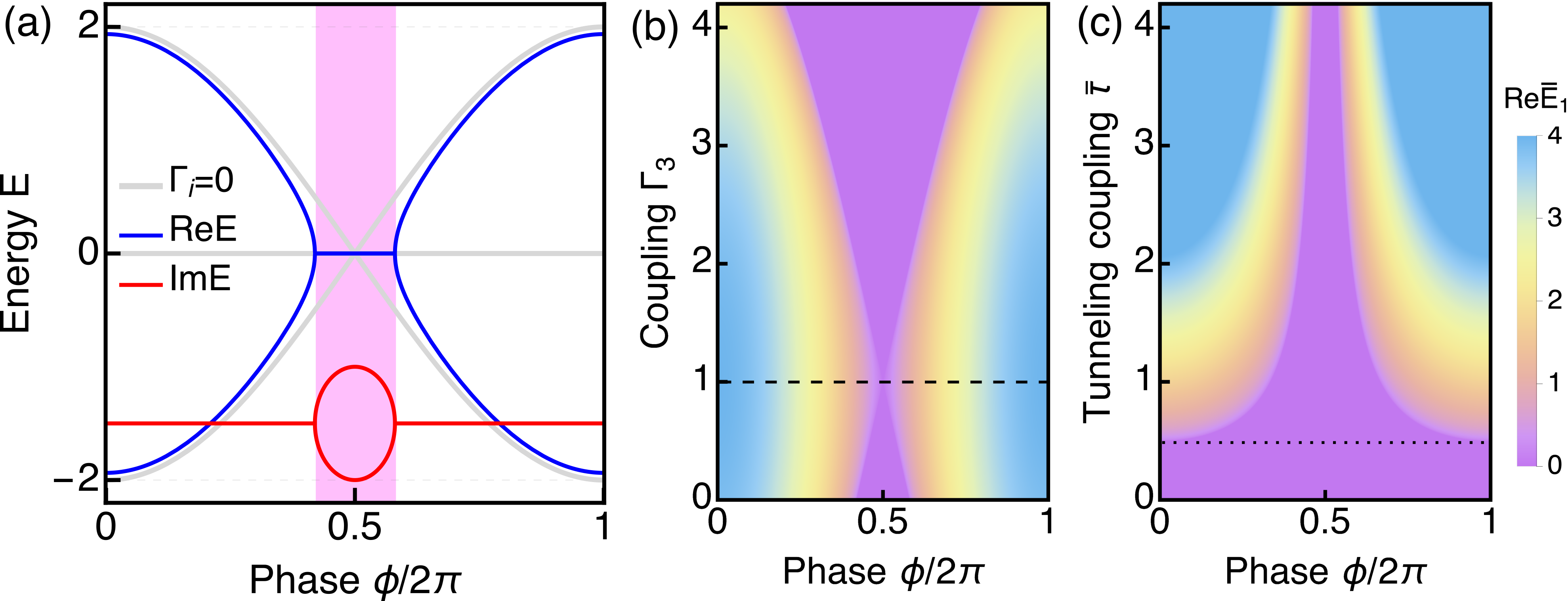}
\caption{(a,b) Re (blue) and Im (red) parts of the eigenvalues as a function of   $\phi$ for $\Gamma_{2}=1$ and $\Gamma_{3}=2$  at $\bar{\tau}=2$, $t'=0$, $t=0$ and $\Gamma_{1,4}=0$. The gray curves correspond to the eigenvalues at $\Gamma_{i}=0$.   (b) Re part of the energy difference between the  first excited  positive and negative levels of (a) as a function of $\phi$ and $\Gamma_{3}$ at $\Gamma_{2}=1$ marked by the horizontal dashed line. (c) The same quantity as in (b) but as a function of $\phi$ and the tunneling coupling $\bar{\tau}$ between superconductors forming the Josephson junction. The ends of the shaded magenta regions in (b,c) mark the position of EPs connecting zero  Re energy lines indicated by magenta color.  The horizontal dotted line in (c) indicates that below it the Re energies are zero for any $\phi$ but no EPs appear.}
\label{Fig4} 
\end{figure}

\subsection{Effect of non-Hermiticity when the outer MZMs are dispersionless in the Hermitian regime}
\label{subsectionIIIC}
So far we have considered values of the couplings between MZMs to be finite, see Eq.\,(\ref{EqHJJ}). In this part, we focus on the regime where $t',t\ll \tau$, which effectively assumes that the outer MZMs are not coupled to the rest and hence possess zero energy in the Hermitian regime, as we show in Fig.\,\ref{Fig1}(b). To inspect the formation of EPs, we calculate the eigenvalues of Eq.\,(\ref{NHJJEq}) and find
\begin{equation}
\label{ABS00}
\begin{split}
\bar{E}^{0}_{\pm}&=-i\Gamma_{1,4}\,,\\
\bar{E}^{1}_{\pm}&=-i\Gamma_{23}\pm2\sqrt{\bar{\tau}^{2}{\rm cos}^{2}(\phi/2)-\gamma_{23}^{2}}\,,
\end{split}
\end{equation}
where the superscript label 0 (1) denotes the lowest (first excited) Andreev level, while
 $\gamma_{23}=(\Gamma_{2}-\Gamma_{3})/2$, $\Gamma_{23}=(\Gamma_{2}+\Gamma_{3})/2$. For EPs, the square root must vanish, namely, $\bar{\tau}^{2}{\rm cos}^{2}(\phi/2)=\gamma_{23}^{2}$, which give us the phases at which EPs appear in $\bar{E}^{1}_{\pm}$: $\phi_{\rm EP}^{\pm}=\pm{\rm arccos}(1-2\gamma_{23}^{2}/\bar{\tau}^{2})+2\pi n$, with $n\in \mathbb{Z}$. Moreover, for $\bar{\tau}^{2}{\rm cos}^{2}(\phi/2)<\gamma_{23}^{2}$, the Re part of the  higher energy ABSs $\bar{E}^{1}_{\pm}$ acquires split Im parts and develops a zero Re energy line connecting the formed EPs. An interesting property of Eqs.\,(\ref{ABS00}) is that for EPs to appear due to $\bar{E}^{1}_{\pm}$, there has to be a distinct amount of non-Hermiticity applied to the inner MZMs characterized by $\gamma_{23}\neq0$, unlike the cases studied in the previous subsection. This is a direct consequence of the outer MZMs being dispersionless already in the Hermitian regime due to $t,t'\ll \tau$ in this case, see Fig.\,\ref{Fig1}(b). To visualize the formation of EPs due to the higher energy ABSs given by Eqs.\,(\ref{ABS00}),  in Fig.\,\ref{Fig4}(a)  we plot  $\bar{E}^{1}_{\pm}$    as a function of $\phi$ at $\Gamma_{2}\neq\Gamma_{3}$. As anticipated, EPs emerge around $\pi$ connected by a zero Re energy line, see shaded region in Fig.\,\ref{Fig4}(a). The Im parts are degenerated below and above EPs   with an energy of $-\gamma_{23}$, while between EPs such Im energies are split and exhibit  a circular dependence around $\phi=\pi$; this dependence is different from the egg-shape profile seen in Fig.\,\ref{Fig3} for finite $t,t'$. We have verified that the wavefunctions corresponding to  $\bar{E}^{1}_{\pm}$ merge into a single wavefunction at EPs, consistent with the definition of EPs \cite{Heiss_2012,OS23,doi:10.1080/00018732.2021.1876991,RevModPhys.93.015005}. The control and tunnability of EPs connected by zero Re energy lines is further supported by Fig.\,\ref{Fig4}(b), which shows the Re part of the difference between $\bar{E}^{1}_{\pm}$ as a function of $\phi$ and $\Gamma_{3}$ at $\Gamma_{1}\neq0$.  Furthermore, from the expressions of  $\bar{E}^{1}_{\pm}$ in Eqs.\,(\ref{ABS00}), one also notes that the condition for EPs can be also controlled by the tunneling coupling between superconductors $\bar{\tau}$.  To see this, Fig.\,\ref{Fig4}(c) shows the   Re part of the difference between $\bar{E}^{1}_{\pm}$ as a function of $\phi$ and $\bar{\tau}$, revealing the formation of EPs and zero energy lines at the edges of the magenta region. It is important to remark, however, that below the dotted line in Fig.\,\ref{Fig4}(c), the ABSs $\bar{E}^{1}_{\pm}$ become flat at zero Re energy but without EPs, see also Eq.\,(\ref{ABS00}). To close, we have demonstrated the formation of  EPs and stable zero Re energy lines as a result of only having inner MZMs subjected to distinct amounts of non-Hermiticity and absent outer MZMs.

%%%%%%%%%%%%%%%%%%%%%%%%%%%%%%%
% SECTION 4:                  Spectral signatures                       %
%%%%%%%%%%%%%%%%%%%%%%%%%%%%%%%
\begin{figure*}[!t]
\centering
\includegraphics[width=0.85\textwidth]{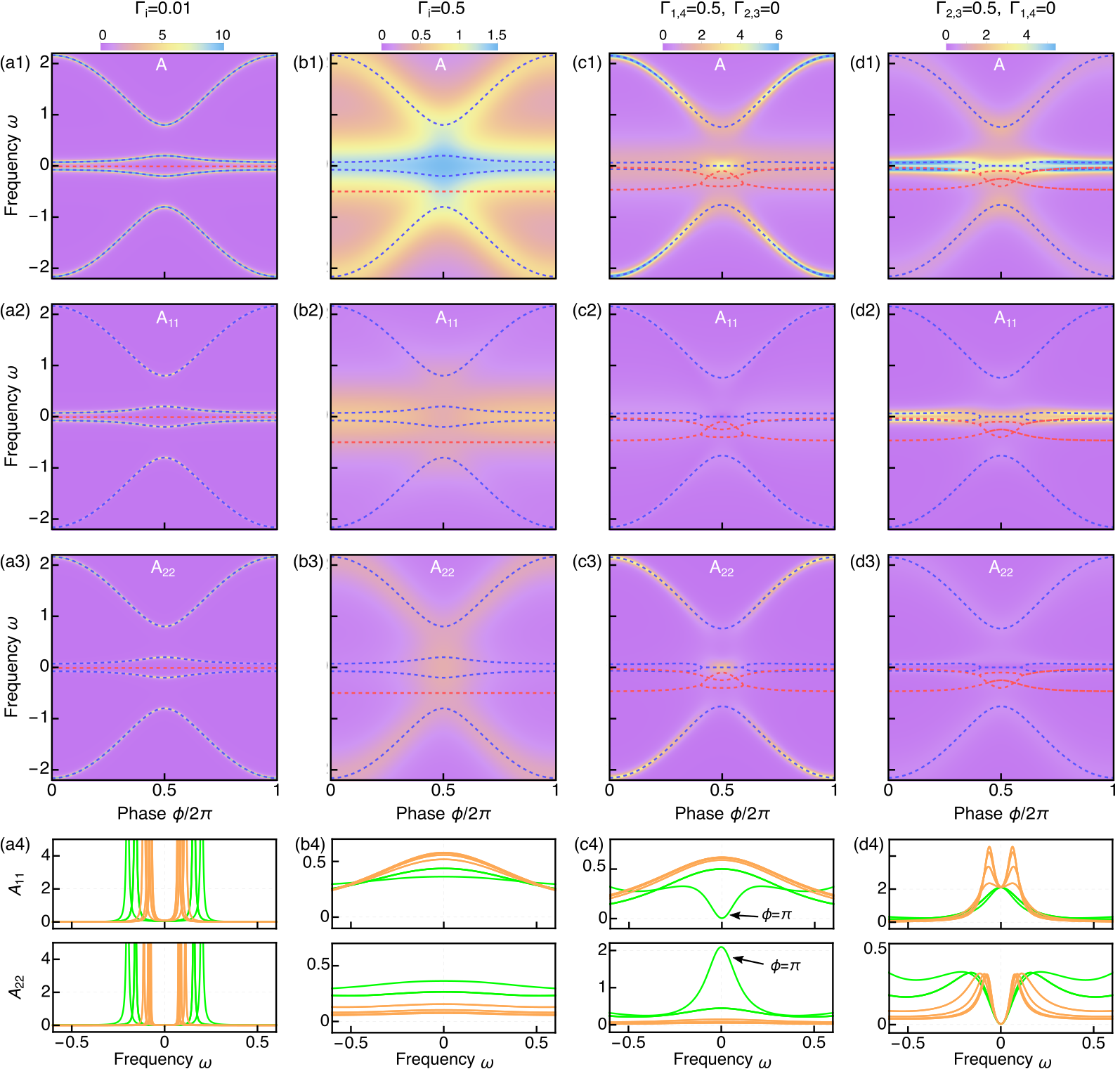}
\caption{Spectral function as a function of frequency $\omega$ and phase difference $\phi$ for distinct values of the coupling to the normal leads $\Gamma_{i}$. Top first row:  Total spectral function at weak equal couplings $\Gamma_{i}=0.01$ (a1), at  sizeable equal couplings $\Gamma_{i}=0.5$ (b1), at  $\Gamma_{1,4}=0.5$ and $\Gamma_{2,3}=0$ (c1), at  $\Gamma_{2,3}=0.5$ and $\Gamma_{1,4}=0$ (d1). Second and third rows: the same as for the top row but the spectral function at the position of the first and second MZMs, and depicted by $A_{11}$ and $A_{22}$,  respectively. Fourth row: line cuts of  $A_{11}$ and $A_{22}$ as a function of $\omega$ for $\phi\in[0,2\pi]$. Here, the green curves correspond to $\phi\in[0.8\pi,1.2\pi]$,  spanning all the $\phi$ values giving rise to zero real energy lines with the ends being the EPs in (c4,d4). The orange curves correspond to $\phi\notin[0.8\pi,1.2\pi]$. In the first, second, and third rows, the dashed blue and red curves show the Re and Im eigenvalues.  Parameters:  $\bar{\tau}=2$, $t'=0.3$, $t=0.5$, $\eta=0.001$.
}
\label{Fig5} 
\end{figure*}

\section{Spectral signatures of exceptional points}
\label{section4}
Having demonstrated the emergence of EPs and stable zero energy states due to MZMs under non-Hermiticity, this section addresses their local and nonlocal spectral signatures. In particular, we inspect spectral properties of the Green's function associated with the effective Hamiltonian [Eq.\,(\ref{NHJJEq})] and found as 
$G_{\rm eff}(\omega,\phi)=-\langle\mathcal{T}\psi\psi^{\dagger}\rangle=(\omega-H_{\rm eff}(\phi))^{-1}$ where $\psi=(\gamma_{1},\gamma_{2},\gamma_{3},\gamma_{4})$ and $\mathcal{T}$ is the time-ordering operator. Thus, the Green's function $G_{\rm eff}(\omega,\phi)$ is a matrix with elements denoted as $G_{nm}(\omega,\phi)$ and allow us to quantify the spectral signatures of local ($n=m$) and nonlocal ($n\neq m$) correlations. The total spectral function is found as $A(\omega,\phi)=-{\rm Im Tr}[G_{\rm eff}(\omega+i\eta,\phi)]\equiv-{\rm Im}[\sum_{n=1}^{4}G_{nn}(\omega+i\eta,\phi)]$, with 
 \begin{equation}
  \label{localGnn}
 \begin{split}
  G_{11}(\omega,\phi)&=\frac{K_{11}-\bar{\tau}^{2}(\omega+i\Gamma_{4}){\rm cos}^{2}(\phi/2)}{\tilde{D}(\omega,\phi)}\,,\\
   G_{22}(\omega,\phi)&=\frac{K_{22}}{\tilde{D}(\omega,\phi)}\,,\\
     G_{33}(\omega,\phi)&=\frac{K_{33}}{\tilde{D}(\omega,\phi)}\,,\\
   G_{44}(\omega,\phi)&=\frac{K_{44}-\bar{\tau}^{2}(\omega+i\Gamma_{1}){\rm cos}^{2}(\phi/2)}{\tilde{D}(\omega,\phi)}\,,
 \end{split}
 \end{equation}
 where $K_{11}=-t^{2}(\omega+i\Gamma_{2})-t'^{2}(\omega+i\Gamma_{3})+(\omega+i\Gamma_{2})(\omega+i\Gamma_{3})(\omega+i\Gamma_{4})$, $K_{22}=-t^{2}(\omega+i\Gamma_{1})-t'^{2}(\omega+i\Gamma_{4})+(\omega+i\Gamma_{1})(\omega+i\Gamma_{3})(\omega+i\Gamma_{4})$, $K_{33}=K_{22}(\Gamma_{1}\leftrightarrow \Gamma_{4},\Gamma_{2}\leftrightarrow \Gamma_{3})$, $K_{44}=K_{11}(\Gamma_{1}\leftrightarrow \Gamma_{4},\Gamma_{2}\leftrightarrow \Gamma_{3})$. Moreover,  
 \begin{equation}
 \label{tildeD}
 \tilde{D}(\omega,\phi)=\omega^{2}\tilde{\mathcal{P}}(\omega,\phi)+2i\omega\tilde{\Gamma}\,\tilde{\mathcal{P}}(\omega,\phi)+\tilde{\mathcal{R}}(\omega,\phi)\,,
 \end{equation}
 with 
 \begin{equation}
 \label{tildePQR}
 \begin{split}
 \tilde{\mathcal{P}}(\omega,\phi)&=\omega^{2}-2(t'^{2}+t^{2})-f(\Gamma_{i})\,,\\
 \tilde{\mathcal{Q}}(\omega,\phi)&=\omega^{2}-t'^{2}-t^{2}-g(\Gamma_{i})\\
 &-[\Gamma_{14}/\tilde{\Gamma}]\bar{\tau}^{2}{\rm cos}^{2}(\phi/2)\,,\\
  \tilde{\mathcal{R}}(\omega,\phi)&=(t'^{2}-t^{2})^{2}+t^{2}(\Gamma_{1}\Gamma_{2}+\Gamma_{3}\Gamma_{4})\\
  &+t'^{2}(\Gamma_{1}\Gamma_{3}+\Gamma_{2}\Gamma_{4})+\Gamma_{1}\Gamma_{2}\Gamma_{3}\Gamma_{4}\\
  &+\Gamma_{1}\Gamma_{4}\bar{\tau}^{2}{\rm cos}^{2}(\phi/2)\,.
 \end{split}
 \end{equation}
and $\tilde{\Gamma}=(\Gamma_{1}+\Gamma_{2}+\Gamma_{3}+\Gamma_{4})/2$, $\Gamma_{nm}=(\Gamma_{n}+\Gamma_{m})/2$, $f(\Gamma_{i})=2(\Gamma_{1}\Gamma_{23}-\Gamma_{2}\Gamma_{34}-\Gamma_{4}\Gamma_{13})$, $g(\Gamma_{i})=[\Gamma_{2}\Gamma_{3}\Gamma_{14}+\Gamma_{1}\Gamma_{4}\Gamma_{23}]/\tilde{\Gamma}$. Note that Eqs.\,(\ref{tildeD}), (\ref{tildePQR}) reduce to Eqs.\,(\ref{barD}), (\ref{barPQR}) and Eqs.\,(\ref{DDOuter14}), (\ref{DDOuter14X}) under the appropriate conditions for $\Gamma_{i}$. Using Eqs.\,(\ref{localGnn}), the spectral function at the position of the $n$-th MZM can be obtained as $A_{nn}(\omega,\phi)=-{\rm Im}[G_{nn}(\omega+i\eta,\phi)]$.  Since $A_{nn}$ can be accessed by local conductance measurements \cite{datta1997electronic}, it is useful to inspect its behavior in order to identify the EPs and the zero  Re energy lines discussed in the previous section. For this purpose, we analyse  $A_{11}$ and $A_{22}$ as well as the total spectral function $A$, which is presented in Fig.\,\ref{Fig5} as a function of $\phi$ and $\omega$ for distinct values of $\Gamma_{i}$ and $t,t'\neq0$. In Fig.\,\ref{Fig5}(a4-d4) we also show $A_{11,22}$ as a function of $\omega$ for phases around $\phi=\pi$.

For vanishing small amounts of non-Hermiticity $\Gamma_{i}=0.01$, the total and local spectral functions directly measure the formation of ABSs, see  Fig.\,\ref{Fig5}(a1-a4) and the overlaid blue and red dashed curves; see also Fig.\,\ref{Fig1}(a).  While the ABSs are captured by both $A_{11}$ and $A_{22}$, the ABSs formed by  outer MZMs and appearing closest to $\omega=0$  produce a stronger intensity in $A_{11}$ for all $\phi$ [Fig.\,\ref{Fig5}(a1,a2)]; similarly, it is $A_{22}$  the spectral function developing a stronger intensity due to ABSs formed by inner MZMs, which emerge at higher frequencies [Fig.\,\ref{Fig5}(a1,a3)]. At $\phi=\pi$, the energy splitting between the four MZMs gives rise to a gapped spectral function with the energy of the outer MZMs defining the gap around $\omega=0$, see Fig.\,\ref{Fig5}(a4). When the amount of non-Hermiticity on the four MZMs is nonzero and equal, the spectral functions do not possess crystal clear signatures to directly identify ABSs due to the large induced broadening, see  Fig.\,\ref{Fig5}(b1-b4) for $\Gamma_{i}=0.5$. In this case, the ABSs due to outer MZMs (closest to $\omega=0$) produce the largest intensity in the total spectral function, although the signal  from inner MZMs also takes sizeable total spectral values coming from $A_{22}$, see Fig.\,\ref{Fig5}(b1-b3). Nevertheless, the intensities are too broad to identify the exact ABSs, as also shown in Fig.\,\ref{Fig5}(b4)
for the local spectral functions around $\phi=\pi$.

When non-Hermiticity applied to the outer MZMs is different from that applied to the inner MZMs, the spectral functions develop interesting features related to the formation of EPs and their zero Re energy lines. This is demonstrated in Figs.\,\ref{Fig5}(c1-c4) and  Fig.\,\ref{Fig5}(d1-d4) for $\Gamma_{1,4}=0.5$, $\Gamma_{2,3}=0$ and $\Gamma_{2,3}=0.5$, $\Gamma_{1,4}=0$, respectively. In the case of outer MZMs under non-Hermiticity, the total spectral function develops strong spectral weight at $\omega=0$ around $\phi=\pi$ due to the zero Re energy line and split Im parts formed between EPs, see Figs.\,\ref{Fig5}(c1). This strong contribution comes from $A_{22}$, which develops a zero-energy peak $\phi=\pi$ while $A_{11}$ has a deep [Figs.\,\ref{Fig5}(c2-c4)]. Moreover, the ABSs formed due to inner MZMs produce  spectral signals that are stronger in $A_{22}$ than in $A_{11}$, suggesting that it is more precise  to access ABS via local conductance in the middle of the JJ when outer MZMs are under non-Hermiticity, see  Figs.\,\ref{Fig5}(c2,c3). The situation is similar when non-Hermiticity is applied to the inner MZMs, where   ABSs forming EPs give rise to strong total spectral weights, thus facilitating their identification, see Fig.\,\ref{Fig5}(d1). 
These strong values come from the local spectral function $A_{11}$, which occurs because non-Hermiticity is applied to the inner MZMs, see Figs.\,\ref{Fig5}(d2,d3). When looking closer at phases around $\phi=\pi$, the local spectral functions $A_{11}$ formes a clear peaked profile around $\phi=\pi$, while $A_{22}$ acquires vanishing values following a V-shape profile [ Figs.\,\ref{Fig5}(d4)]. In this case, when inner MZMs are under non-Hermiticity, identifying Andreev EP signatures would be more accessible by local conductance at the outer sides of the JJ.

\begin{figure}[!t]
\centering
\includegraphics[width=0.48\textwidth]{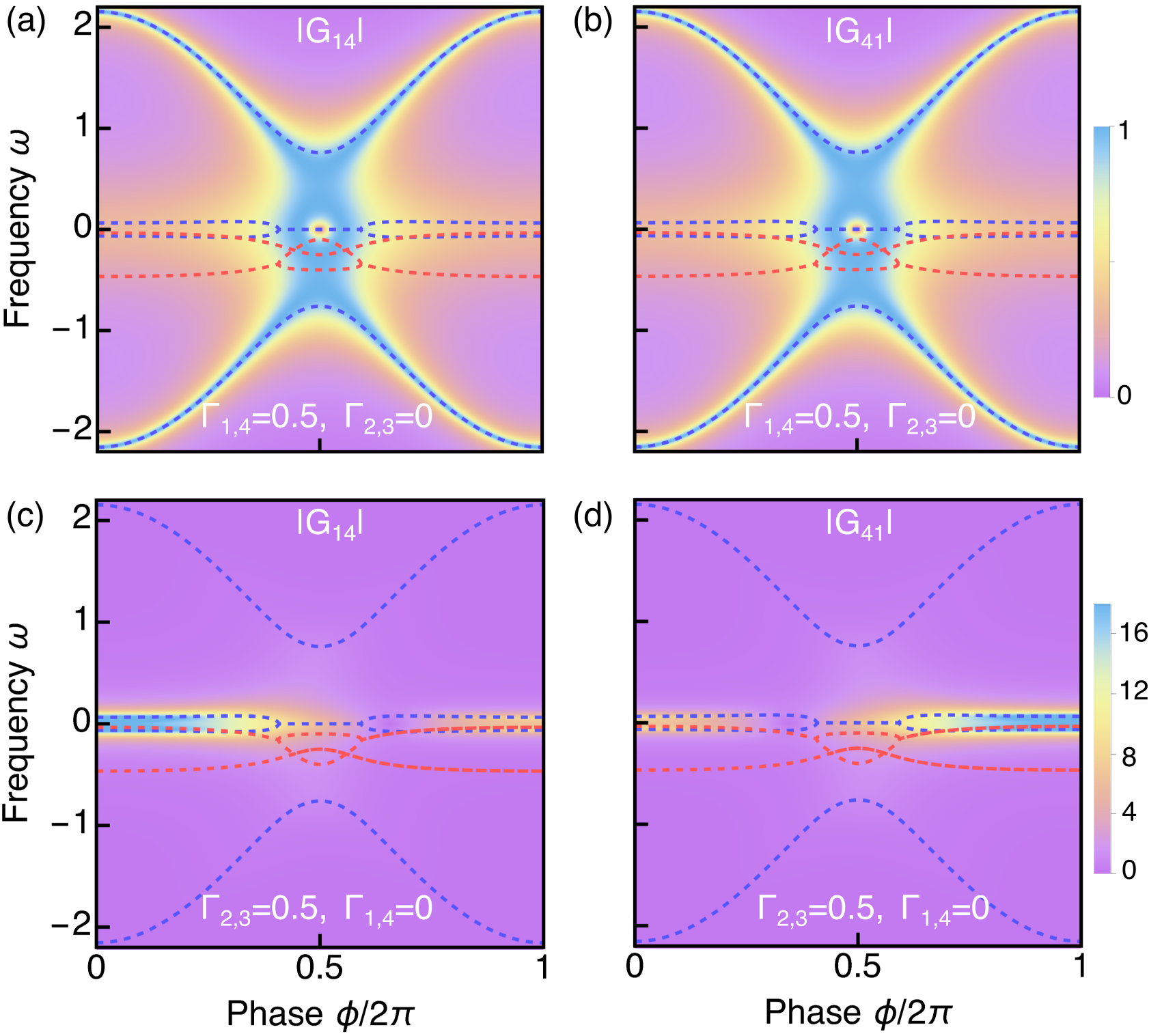}
\caption{Absolute value of the nonlocal Green's function components $|G_{14 (41)}|$ as   functions of $\omega$ and $\phi$ for distinct values of $\Gamma_{j}$: (a,b) $\Gamma_{1,4}=0.5$, $\Gamma_{2,3}=0$, which correspond to the regime of Fig.\,\ref{Fig2}(a); (c,d) $\Gamma_{j}$:  $\Gamma_{2,3}=0.5$, $\Gamma_{1,4}=0$, which correspond to the regime in Fig.\,\ref{Fig3}(a). Parameters:  $\bar{\tau}=2$, $t'=0.3$, $t=0.5$, $\eta=0.001$.
}
\label{Fig6} 
\end{figure}

Besides local spectral signatures, the elements $G_{nm}(\omega,\phi)$ of the effective Green's function $G_{\rm eff}(\omega,\phi)$ also permit us to look at the nonlocal correlations. Of relevance is, for instance, the Green's function element between the outer MZMs ($G_{14}$ and $G_{41}$) as it holds relevance in nonlocal conductance signals \cite{PhysRevB.92.100507,PhysRevB.93.201402,Cayao2024Controllable,cayaominimalkitaev}. To be more precise, we find that $G_{14}$ and $G_{41}$ are given by
\begin{equation}
\label{EqG1441}
\begin{split}
   G_{14}(\omega,\phi)&=\frac{2t't(\omega+i\Gamma_{23})+i\bar{\tau}(t^{2}-t'^{2}){\rm cos}(\phi/2)}{\tilde{D}(\omega,\phi)}\,,\\
    G_{41}(\omega,\phi)&=\frac{2t't(\omega+i\Gamma_{23})-i\bar{\tau}(t^{2}-t'^{2}){\rm cos}(\phi/2)}{\tilde{D}(\omega,\phi)}\,,\\
 \end{split}
 \end{equation}
 where $\Gamma_{23}=(\Gamma_{2}+\Gamma_{3})/2$ and $\tilde{D}(\omega,\phi)$ is given by Eq.\,(\ref{tildeD}). The first property we see in these nonlocal Green's functions is that, besides sharing the same denominator with the local components, their numerator is influenced by non-Hermiticity via the couplings between MZMs ($t,t'$). The phase dependent term in the numerator is, however, not affected by non-Hermiticity. Interestingly, these functional dependences have consequences that allow to identify the formation of EPs, as visualized in Fig.\,\ref{Fig6} where we plot the absolute value of $G_{14,41}$ as a function of $\phi$ and $\omega$ for non-Hermiticity applied to inner and outer MZMs, while maintaining  $t,t'\neq0$. For non-Hermiticity applied only to outer MZMs ($\Gamma_{1,4}\neq0$ and $\Gamma_{2,3}=0$), the nonlocal correlations  $G_{14,41}$ acquires similar large values between the EPs formed by the   ABSs closest to zero, and also pick up large intensities at the frequencies of the higher energy ABSs, see Fig.\,\ref{Fig6}(a,b). While it is relatively simple to mark the position of the formed EPs by overlaying the energies on the nonlocal correlation, the sizeable Im parts produce broad signals around $\omega=0$ that makes it difficult to clearly position the Re ABSs [Fig.\,\ref{Fig6}(a,b)]. In the case of having non-Hermiticity applied only to the inner MZMs [Fig.\,\ref{Fig6}(c,d)], the nonlocal correlations  $G_{14}$ and  $G_{41}$ develop large intensities below and above $\phi=\pi$, respectively. This effect is tied to the numerators of  $G_{14,41}$ in 
 Eq.\,(\ref{EqG1441}), and the length of these strong intensity regions along $\phi$ makes it possible to identify the formation of EPs, see Fig.\,\ref{Fig6}(c,d). As a result, EPs and their zero Re energy lines can be revealed by nonlocal spectral signatures.

%%%%%%%%%%%%%%%%%%%%%%%%%%%%%%%
% SECTION 5:                       CONCLUSIONS                       %
%%%%%%%%%%%%%%%%%%%%%%%%%%%%%%%
\section{Conclusions}
\label{section5}
In conclusion, we have investigated  phase-biased Josephson junctions with four Majorana zero modes under the effect of non-Hermiticity due to their coupling to normal leads. We have demonstrated   that  these non-Hermitian Josephson junctions host second order exceptional points    entirely controlled by the interplay between the Josephson effect and an asymmetric amount of non-Hermiticity. In particular, we have shown that a finite non-Hermiticity applied either on the inner or outer Majorana zero modes is sufficient to realize stable exceptional points between the lowest Andreev bound states, while a stronger non-Hermiticity is able to further induce exceptional points   between the first excited Andreev states. As a result, these Andreev exceptional points form even when there exist a finite overlap between Majorana wavefunctions. Interestingly, the formation of  Andreev exceptional points also enables the emergence of zero real energy lines connecting them, akin to bulk Fermi arcs \cite{doi:10.7566/JPSCP.30.011098,PhysRevB.107.104515,PhysRevB.97.041203,PhysRevB.99.165145,kozii2017} but here driven by the effect of non-Hermiticity on the Josephson effect. This makes non-Hermiticity a powerful way for manipulating Andreev bound states in Josephson junctions.   We have further shown that the obtained Andreev exceptional points and their zero-energy bands produce measurable local and nonlocal spectral signatures, which can be exploited for detecting the reported non-Hermitian Josephson phenomena.  We anticipate that the obtained Andreev exceptional points are expected to be robust against  sources of disorder that preserve the underlying particle-hole symmetry of class $D^{\dagger}$ \cite{PhysRevX.9.041015,PhysRevLett.123.066405}. Since exceptional points are directly tied to non-Hermitian topology \cite{OS23,PhysRevX.9.041015,PhysRevLett.123.066405},   our findings  unveil the realization of topologically protected  phenomena controlled by the Josephson effect and non-Hermiticity.

%%%%%%%%%%%%%%%%%%%%%%%%%%%%%%%
%                        ACKNOWLEDGMENTS                               %
%%%%%%%%%%%%%%%%%%%%%%%%%%%%%%%

\section{Acknowledgements}
J. C. acknowledges  financial support from the Swedish Research Council  (Vetenskapsr\aa det Grant No.~2021-04121) and the Carl Trygger’s Foundation (Grant No. 22: 2093).  
M.S. are supported by JST CREST (Grant No. JPMJCR19T2) and JSPS KAKENHI (Grant Nos. JP24K00569 and JP25H01250).
The computations  were  enabled by resources provided by the National Academic Infrastructure for Supercomputing in Sweden (NAISS), partially funded by the Swedish Research Council through grant agreement no. 2022-06725.

%%%%%%%%%%%%%%%%%%%%%%%%%%%%%%%
%                                  REFERENCES                                  %
%%%%%%%%%%%%%%%%%%%%%%%%%%%%%%%

\bibliography{biblio}

\end{document}